\documentclass[%
 aip,
amsmath,
amssymb,
reprint,%
]{revtex4-1}
\usepackage{graphicx}
\usepackage{dcolumn}
\usepackage{bm}

\usepackage[utf8]{inputenc}
\usepackage[T1]{fontenc}
\usepackage{mathptmx}
\usepackage{etoolbox}
\usepackage{lineno}
\usepackage{xcolor}
\usepackage{amsmath}
\usepackage{physics}

\makeatletter
\def\@email#1#2{%
 \endgroup
 \patchcmd{\titleblock@produce}
  {\frontmatter@RRAPformat}
  {\frontmatter@RRAPformat{\produce@RRAP{*#1\href{mailto:#2}{#2}}}\frontmatter@RRAPformat}
  {}{}
}%
\makeatother

\begin{document}


\title{A nonlinear room mode determines the operating conditions of a large-cavity synthetic jet actuator at low frequencies}

\author{L.F. Olivera-Reyes}
\affiliation{Instituto de Ciencias Aplicadas y Tecnolog\'ia, Universidad Nacional Aut\'onoma de M\'exico, Circuito Exterior S/N, Ciudad Universitaria, 04510, Mexico City.}

\author{E.S. Palacios de Paz}
\affiliation{Instituto de Ciencias Aplicadas y Tecnolog\'ia, Universidad Nacional Aut\'onoma de M\'exico, Circuito Exterior S/N, Ciudad Universitaria, 04510, Mexico City.}

\author{S. S\'anchez}
\affiliation{Instituto de Ciencias Aplicadas y Tecnolog\'ia, Universidad Nacional Aut\'onoma de M\'exico, Circuito Exterior S/N, Ciudad Universitaria, 04510, Mexico City.}

\author{J.F. Hern\'andez-S\'anchez}
\affiliation{Instituto de Ciencias Aplicadas y Tecnolog\'ia, Universidad Nacional Aut\'onoma de M\'exico, Circuito Exterior S/N, Ciudad Universitaria, 04510, Mexico City.}
\email{federico.hernandez@icat.unam.mx}

\date{\today}

\begin{abstract}
Synthetic Jet (SJ) actuators are an intrinsically complex combination of electronics, electric and mechanical systems. When studied theoretically, these elements are often simplified to coupled damped harmonic oscillators (DHO) that induce a pressure field within the cavity and drive momentum exchange. Thus, the performance of an SJ actuator results from coupling these DHOs, naturally leading to a few resonant modes. There is good evidence in the specialized literature of two resonant modes developing in SJ actuators: the membrane/piezoelectric mode and the Helmholtz resonance. In this work, we report on the effect of a third resonant mode that develops at very low frequencies due to a cavity much larger than the volume displaced by the actuator. We present evidence that the large-cavity dynamics determine the SJ performance in combination with the well-described formation criteria. We compare the intensity of this resonant mode with the first room modes using standard frequency analysis. Unlike typical room modes, the distribution of this resonant mode is very biased to lower frequencies. We also show that the resonant mode may be dimmed and focused by adding an obstacle in different cavity positions for the lower sound intensities. This mode overcomes the Helmholtz resonance, dominating the dynamics for higher sound intensities. We show that jet and vortex velocities mimic the sound pressure curve for the low-frequency range. Its effect mitigates for the higher range due to a delve through smaller stroke lengths, characterized as a fixed relation between the Reynolds and the Stokes numbers. We consider that the large-cavity dynamics is an additional element that, if integrated as design criteria, could extend the optimum frequency response of SJs.
\end{abstract}

\maketitle



\section{Introduction}

Synthetic jet (SJ) actuators produce a periodic flow outwards the neck of a Helmholtz resonator operating as intermittent pumps. Since the cavity of the resonator refills during the negative oscillation cycle, the outward flow does not require the addition of net mass\cite{Mohseni2014}. Thus, SJs are alternatively known in the literature as zero-net-mass-flux (ZNMF) jets. Despite not adding net mass, SJs produce periodic bursts of momentum \cite{Krishnan2009}. The SJ dynamics have been investigated extensively for more than 20 years now\cite{Glezer2002}.  One of the most important findings was the
vortex formation criteria in dimensionless parameters\cite{Smith1998, Cater2002, Brouvckova2016}. The two most relevant dimensionless parameters that summarize the vortex formation dynamics are the Reynolds and the Stokes numbers. In this chart, curves of Re $\sim$ S$^2$ correspond to constant dimensionless stroke lengths. In this relationship, a prefactor of 0.16 corresponds to a dimensionless stroke length of 0.25 and the threshold of the vortex formation criteria. Even though the dimensionless stroke length is critical for the vortex formation, the transition through different prefactors influences the velocity and flow rate. Thus, crossing through dimensionless stroke lengths does not represent a sharp transition but changes to weaker flow rates. 

A typical SJ actuator combines two components: (1) a transducer and (2) a Helmholtz resonator\cite{Chiatto2017}. The transducer may be an electroacoustic (speaker) or a piezoelectric membrane. Typically, the function of the transducer is to change the volume of the cavity periodically. In some cases, the flow is considered incompressible\cite{Travnivcek2015}, but in others the neck of the resonator limits the outflow, leading to pressure accumulation in the cavity\cite{Amitay2006}. Through the lens of vibration theory, a speaker has equivalent spring and damping constants, and the cone and the dust cap have inertia. Conversely, the air within the cavity of the Helmholtz resonator acts as a spring, and the fluid on its neck acts as a mass\cite{Rayleigh1916}. Here, the fluid friction with the neck transforms the viscosity and the nozzle geometry to a damping constant. Both the transducer and the resonator have elements of damped harmonic oscillators (DHOs). In standard vibration theory the mass, the spring, and damping constants are simplified to the resonant frequency and the damping ratio. Changes in the resonant frequency have been studied along the operational range of SJ actuators. For example, SJ actuators exhibit at least one resonant frequency\cite{Chiatto2017,Ahmed2009, Alvi2010}. But since there are at least two elements that can be simplified as DHOs, some SJ actuators have exhibited two resonant frequencies\cite{Sharma2007,Gallas2003}. 

\begin{table*}[t]
\centering
\caption{\label{Tab:SJDimensions}{Cavity maximum dimensions (diameter and length), and maximum frequency of operation of different SJ actuators in the literature.}}
\mbox{}\\
\begin{tabular}{|l|c|c|c|c|} \hline
Reference & $D_\textrm{max}$ [mm] & $L_\textrm{max}$ [mm] & $V_\textrm{max}$ [mm$^3$]& $f_{max}$ [Hz]  \\ \hline
Chiatto \it{et al.}\cite{Chiatto2018}  (2018)			&   6 &  6    & 1.7$\times$10$^2$ &3000\\
Krishnan \& Mohseni\cite{Krishnan2009}  (2009)	&  40&  3.4 & 4.3$\times$10$^3$ &3000\\
Chiatto \it{et al.}\cite{Chiatto2019} (2019)			& 42	&  7.5 & 1.0$\times$10$^4$ &3000\\
Kord\'ik \& Tr\'avn\'i\u{c}ek\cite{Kordik2017} (2017)	& 60 & 10   & 2.8$\times$10$^4$ & 170 \\
Kord\'ik \& Tr\'avn\'i\u{c}ek\cite{Kordik2018} (2018)	&60.8& 10  & 2.9$\times$10$^4$ & 150 \\
F\"orner \& Polifke\cite{Forner2015} (2015)		&50.8&25.4& 5.1$\times$10$^4$ &1000\\
Persoons \& O'Donovan\cite{Persoons2007} (2007)	& 44  &  75 & 1.1$\times$10$^5$ &1000\\
Gil \& Strzelczyk\cite{Gil2016} (2016)			& 50	 & 60  & 1.2$\times$10$^5$ & 600\\
Ingard \& Ising\cite{Ingard1967}  (1967)			& 95  &  70 & 5.0$\times$10$^5$ &  370\\
Smyk et al.\cite{Smyk2020} (2020)				& 164 & 30 & 6.3$\times$10$^5$ & 200\\
This study									& 127 &200 &8.3$\times$10$^6$ &300\\ \hline
\end{tabular}
\end{table*}

Lumped Element Models, a common strategy to study SJ actuators, collapse the three dimensions of these devices to a single point, which is only possible if the pressure and velocity fields are accurately portrayed by their averaged values\cite{Chiatto2018}. For a speaker, this assumption signifies that the cone displacement occurs in a single direction. Although this is not true in some vibration modes\cite{ Chiatto2017, Chiang2015}, this assumption is valid if the dominant length scales are much larger than the device dimensions. For an SJ, the dominant length scale is the sound wavelength. Typical driving frequencies for SJ actuators are on the order of 1000 Hz. If we assume the speed of sound is 340 m/s, then the wavelength is about 300 mm. Most cavities are considerably smaller than this length, as seen in Table~\ref{Tab:SJDimensions}. The dimensions of SJ actuators, reported in Table~\ref{Tab:SJDimensions}, are for several applications, ranging from efficiency characterization\cite{Gil2016} to plasma ejection\cite{Chiatto2018}. Other applications include the study of the flow field driven by a piezoelectric actuator\cite{Krishnan2009}, and the coupling of formation criteria with the lumped element models\cite{Chiatto2019}. Table~\ref{Tab:SJDimensions} also includes a study of acoustic dissipation via numeric simulations in a cylindrical cavity\cite{Forner2015} and its corresponding experimental validation\cite{Persoons2007}. We also added the actuator dimensions of the classic work about acoustic non-linearities of an orifice of Ing{\aa}rd and Ising\cite{Ingard1967} (1967). Table~\ref{Tab:SJDimensions} is organized from small to large cavity volumes and shows that even the largest cavity is more than one order of magnitude smaller than the device used in this study. The cavity length is much larger than that of other devices as well. One reason for the size discrepancy is that SJ actuators are commonly designed for the aerospace industry and the fabrication of small devices\cite{Chiatto2017}. The weight limitation of the industry and the small size have fueled an interest in designing SJ actuators as small as possible, which resulted in compact cavities as a rule of thumb. Thus, not only the assumption needed to simplify the SJ actuators is typically met, but the frequencies required to form room modes within the cavity of SJ actuators are typically higher than the operational range. Thus, Lumped Element (LE) models correctly assumed SJ actuators are small compared with acoustic wavelengths\cite{Crocker1998}. 

In any room in a house or building, some sound frequencies are amplified more than others due to their geometry. In a rectangular, closed, and empty room, the amplified frequencies correspond to the formation of standing waves due to reflection. The fundamental mode develops when the sound wavelength is at least half the distance between parallel walls. The process repeats for higher harmonics. These wavelengths are called room modes and are determined correctly by linear room acoustics\cite{Kuttruff2016}. Standard room acoustics is used extensively to predict the frequencies amplified in rooms where human activities occur, such as concert halls and recording studios. The larger the room, the smaller the frequency amplified in it. In this context, a small room measures at least two meters, and the frequencies are as low as a hundred Hertz\cite{Kleiner2014}. Smaller rooms, such as cavities of SJ actuators, have not been characterized using standard room acoustics because human activities do not occur in them. However, in micro- and nano-particle manipulation, it has been reported that an acoustic field groups particles by size in different cavity locations. Such is the case of Qiao {\it et al.} \cite{Qiao2015} (2015), where aerosol drops are organized by size due to the formation of standing waves within a cavity of similar size, 5.4$\times$10$^6$ mm$^3$. In their investigation, the resonant frequencies corresponded to standard room acoustics modes\cite{Kuttruff2016}. While other studies might be relevant\cite{Reyes2018}, the operating frequencies used in aerosol manipulation are much larger than the hearing threshold.

In this work, we revisited the SJ actuator designed and built by Boullosa \& Ordu\~na-Bustamante\cite{Boullosa2010} (2010) to float polystyrene spheres using airflow. The cavity volume of this actuator is one order of magnitude larger than the largest of other SJ devices, and a non-linear room mode develops for a lower frequency than the first room mode. Unlike linear room modes, this one is asymmetric, reaching its peak during the tenths of Hertz and diminishing during hundreds of Hertz. We performed frequency analysis on each element to compare this asymmetric room mode with the effect of previously reported DHOs. We show that this asymmetric room mode dominates the SJ dynamics above the two resonating frequencies that have been reported and found conclusive evidence that the cavity dynamics introduce a third DHO that determines the performance of the SJ actuator. 


\section{Experimental setup}

\subsection{Materials and instruments}
\begin{figure*}
\centering
\includegraphics[width=1.98\columnwidth]{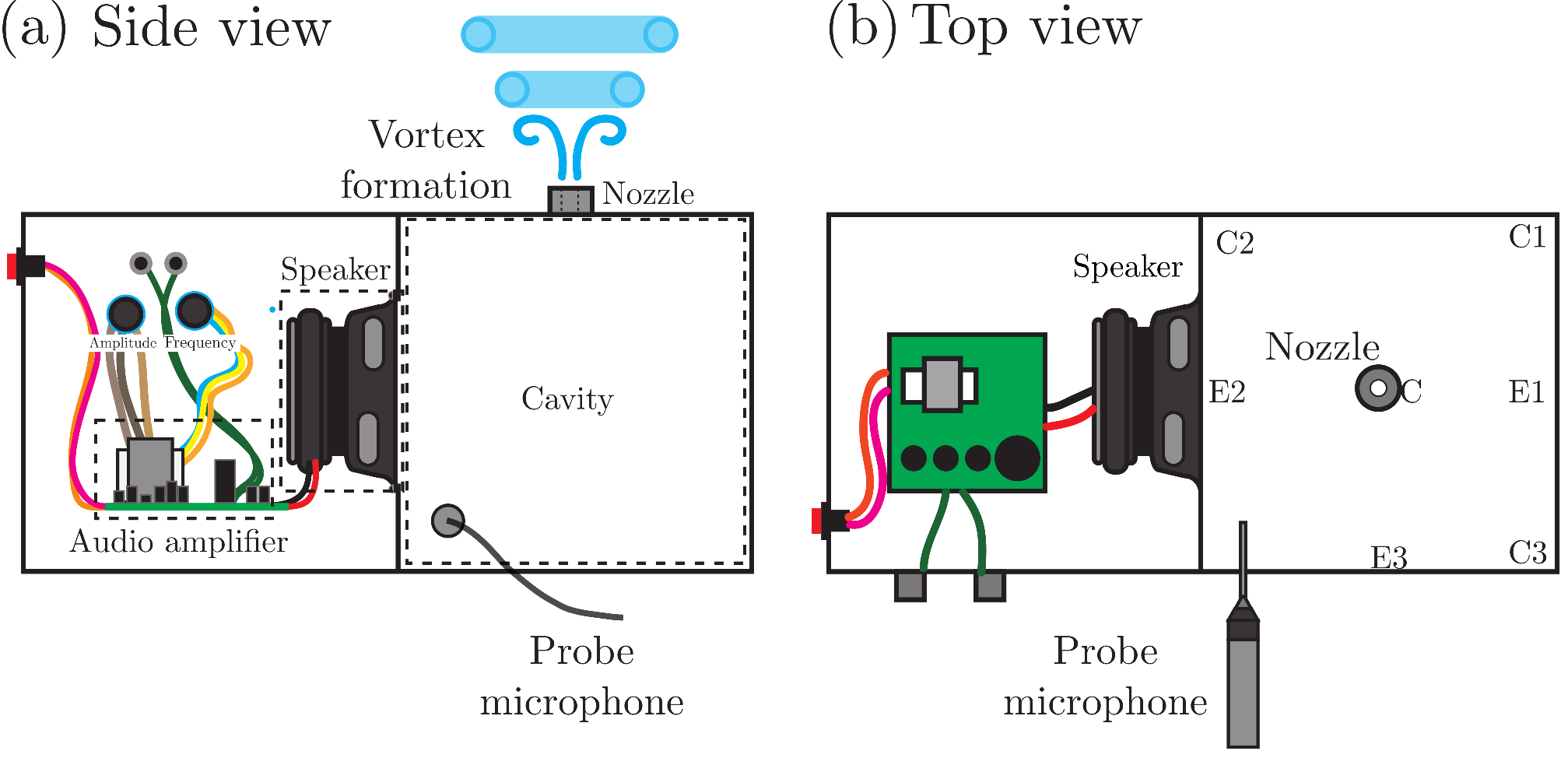}
\caption{(Color online)(a)~Sketch of the SJ actuator used in this work. (a) Side view. The elements of the Synthetic Jet device are highlighted using broken squares. (b)~Top View. The obstacle positions within the cavity.}
\label{Fig:SetupSketch}
\end{figure*}

The SJ actuator, depicted in Figure~\ref{Fig:SetupSketch}, consists of an acrylic box with two inner cavities. The walls have a thickness of 10 mm. The speaker is a Dayton DC160-4 of 6.5'' in diameter, and it is attached to the dividing inner wall. The speaker can be driven at a maximum power of 120 W and has a nominal impedance of 4 $\Omega$. The inner driver consists of an analog sine wave generator and an audio amplifier that drives the speaker at frequencies up to 300 Hz and voltages up to 10 V. On the back of the speaker, there is a closed cavity where the driver is enclosed. The back wall of the cavity is removable and remains sealed in all the experiments presented here. However, we verified the results are the same when the wall is removed. The front cavity has a cubic volume of 20 cm in length. Considering an additional volume due to the recess of the cone, we estimated the cavity volume as 8250 cm$^3$. The front cavity is open to the environment through a nylamid nozzle tightly fitted on top of its top wall. However, in many of our tests, the neck of the SJ actuator remained blocked, decoupling the Helmholtz resonator to study the cavity dynamics independently. For the last set of experiments, we used two nozzles with 8 and 10 mm of diameter, both with a length of 40 mm. Here, we were interested in the effect of coupling two Helmholtz resonators. The inner edges of the neck have a chamfer radius of about 1 mm.

The magnitude of the electrical impedance was measured for low voltages using an Audio Test System (DATS V2 from Dayton Audio). This test system requires the use of a computer and installing the software. We detached the speaker from the SJ actuator to perform this test.

An accelerometer is attached to the speaker dust cap to measure the RMS displacement as a function of the frequency and voltage. The accelerometer attached is an Endevco 27A11 with its signal conditioner Endevco 4416B. The mass of the accelerometer is 1 g. The combined mass of the coil, the cone, and the dust cap is 23 g, as measured in a similar and disassembled speaker. We captured the signal using an oscilloscope (Tektronix MDO3032). The acceleration was calibrated using a vibration calibrator Br\"uel \& Kj\ae r type 4294. This calibrator produces a reference RMS acceleration amplitude of 10 m/s$^2$ and a reference RMS displacement amplitude of 10 $\mu$m at 1000 rad/s ($\approx$159 Hz) of frequency. This calibration resulted in a sensitivity of 10 m/s$^2$ every 102 mV. 

\subsection{Calibration methods and measurements}
Two frequency characterizations were performed in anechoic conditions: (1) the speaker and (2) the cavity responses. For the first test, we placed the speaker in the anechoic chamber of the   \textit{Laboratorio de Ac\'ustica y Vibraciones} of the \textit{Instituto de Ciencias Aplicadas y Tecnolog\'ia}. Then, we performed measurements using a probe microphone at 30 cm above the speaker. As a second test, we fixed the speaker to the SJ actuator and placed the device inside the anechoic chamber. We inserted the microphone-probe tip within the cavity of the SJ actuator through the small hole seen in Figure~\ref{Fig:SetupSketch}. We performed both characterizations by driving the speaker with a sine sweep at constant voltages ranging from 10 mV up to 10 V. This is called frequency analysis in room acoustics. To perform the frequency analysis, we bypassed the inner driver using the digital function generator and the Yamaha audio amplifier to be able to test sine sweeps in a single run. We used a signal analyzer to estimate the maximum sound pressure spectrum. We focused on the pressure measurements when the cavity is blocked, but we also tested 8- and 10-mm nozzles.

The probe microphone used was a Br\"uel \& Kj\ae r type 4182 connected to a power supply Br\"uel \& Kj\ae r type 2807. The signal from the microphone was calibrated before each experimental session using a sound level calibrator Br\"uel \& Kj\ae r type 4230. This calibrator produces a signal at 1000 Hz and an RMS sound pressure of 1 Pa. A digital function generator, Siglent SDG1032X, generated sine sweeps at a constant voltage. The duration of each sweep cycle was 300 seconds. An audio amplifier, Yamaha P2500S (YA), modulated the signal fed to the speaker. We used a multimeter, Fluke 289, to monitor the input voltage. The sound pressure curves were estimated using the maximum peak mode. The signal analyzer was a Stanford Research SR780.

We visualized the external flow using Planar Laser-induced Fluorescence (PLIF). A fog generator (Dantec Dynamics, Rosco Mini-V) attached to the cavity injected smoke through one of the walls via a needle valve. The typical size of the particle tracers is 1 to 3 $\mu$m. A laser sheet, wavelength of 448-462 nm (blue), illuminated a cross-section of the flow downstream of the nozzle. A laser engraver module modified in-house of 20 W produced the laser sheet. A collimator lens focused the laser on a cylindrical lens. The minimum sheet thickness we reached with this arrangement was 1 mm. We designed and built a Pulsed-Width-Modulation (PWM) circuit to drive the laser module at a nominal frequency of 20 kHz. We recorded the experiments using a high-speed camera perpendicular to the flow: Chronos 2.1-HD from Krontech. The typical frame rate was 4229 frames per second (fps). We used a 50 mm CCTV lens and a 30 mm infinity tube to adjust the working distance and record the images. We tracked the front of each vortex in a sequence of recorded frames to estimate its velocity. Using linear regression, we fitted the positions as a function of time and estimated the average velocity of the vortex at about 4 cm downstream of the neck. Such an average relies on the mean position of the vortex through a few frames, being more reliable than instantaneous velocity fields.

In the following section, we report on the speaker frequency response, the cavity, and the effect of two nozzles coupled to the SJ actuator. Then, we compare the behavior of each of these elements to measurements of the speed of the jets and vortices expelled. Finally, in the discussion section, we elaborate arguments that may enlighten the mechanism underlying the dynamics.

\section{Results}
Here, we describe the main elements of the SJ actuator that may significantly affect its dynamics. The speaker is, by itself, a complex device that operates on three different levels: electric, mechanical, and acoustic. In this section, we tested the response of each level independently. The impedance curve illustrates the electric response of the speaker. Curves of the cone RMS displacement as a function of the frequency depict the mechanical response of the speaker. The sound pressure level reports the acoustic response of the speaker. We repeated the measurement within the cavity of the SJ actuator. Then, we measured the velocity of the vortices expelled from the nozzle for different conditions.

\subsection{Speaker characterization}
\begin{figure*}
\centering
\includegraphics[width=1.98\columnwidth]{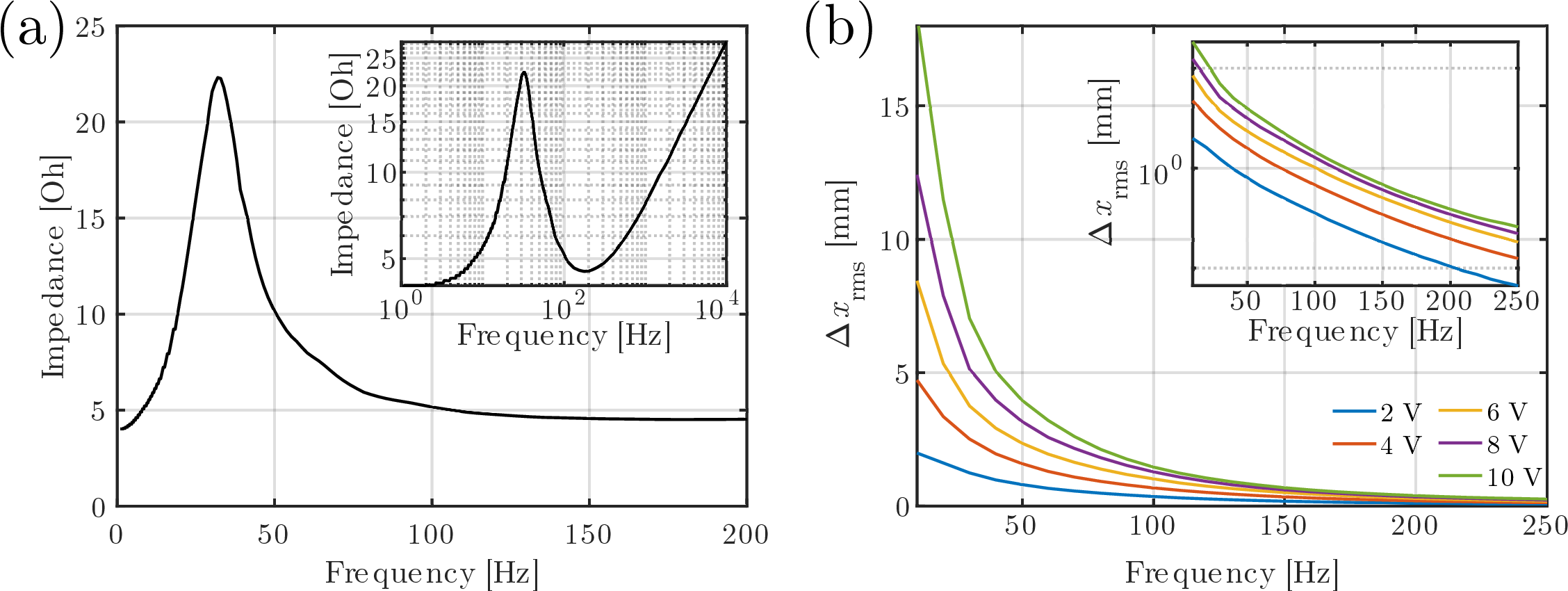}
\caption{(Color online)(a)~Magnitude of the electric impedance as a function of the frequency measured with an Audio Test System. (b)~RMS displacement of the speaker as a function of the frequency for different voltages.}
\label{Fig:Impedance&Displacement}
\end{figure*}

A speaker is a device that converts alternate current to a mechanical vibration through a coil in relative movement with a permanent magnet. The mechanical vibration transforms into sound via the compression of the adjacent air. The coil is associated with an electrical resistance and a reactance that characterizes its electrical response. To estimate the speaker response, we tested it independently by detaching it from the SJ actuator. We used an Audio Test System (DATS V2 from Dayton Audio) to estimate the impedance curve at low voltages and some Thiele-Small parameters. The magnitude of the impedance as a function of the frequency in Figure~\ref{Fig:Impedance&Displacement}(a) shows a spike at 32 Hz. The impedance magnitude reaches its nominal value for lower and higher frequencies, up to 300 Hz. The inset of Figure~\ref{Fig:Impedance&Displacement}(a) shows that the impedance magnitude increases continuously for frequencies higher than 300 Hz. Since the frequency axis of the inset is on a logarithmic scale, the increase in the impedance is gradual.

Here, we investigated if the impedance peak may infiltrate the mechanical or acoustic responses of the speaker. Besides the electrical, the mechanical components could be also conceived as elements of a DHO, \textit{i.e.}, the suspension ring and the spider result in elastic and damping constants, and the coil, the cone, and the dust cap have equivalent mass. Such elements may lead to vibration modes on their own. Thus, we attached the accelerometer to the center of the dust cap to measure its displacement as a function of the driving frequency for different voltages. We performed this test with the speaker fixed to a breadboard in an open room. Figure~\ref{Fig:Impedance&Displacement}(b) shows the Root Mean Square (RMS) displacement as a function of the frequency for different voltages. The inset of Figure~\ref{Fig:Impedance&Displacement}(b) shows the displacement in logarithmic scales. Although the behavior in the main figure seems to fit an Arrhenius-type equation, we observe in the inset that the exponent is not constant. The speaker exhibits larger displacements for lower frequencies. This test reveals that, for lower frequencies, no other DHO is present in the mechanical response, including the magnitude of the electrical impedance. The higher displacements occur on the lower frequency band on the non-audible spectrum.

\begin{figure*}
\centering
\includegraphics[width=1.98\columnwidth]{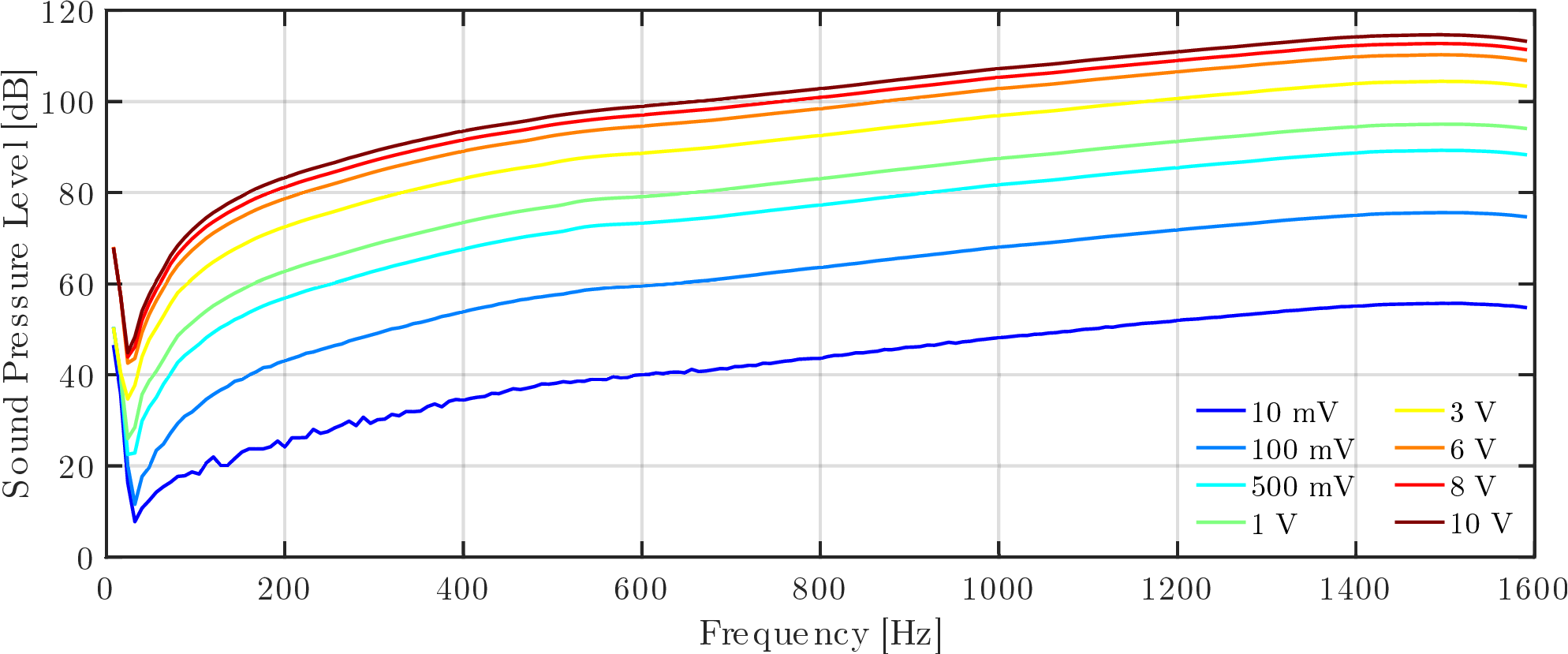}
\caption{(Color online) Speaker frequency response. Maximum sound pressure level during a sine sweep measured in an anechoic chamber from 1 to 1600 Hz for different voltages.}
\label{Fig:SpeakerResponseAlone}
\end{figure*}

To test the speaker's acoustic response, we placed it alone in the anechoic chamber while reproducing the sine sweep and recording with the microphone. Figure~\ref{Fig:SpeakerResponseAlone} shows a sound pressure level average as a function of the frequency for a range of logarithmically spaced voltages. Despite some response below the hearing range, the sound pressure level arises steadily when frequencies reach about 20 Hz. The curve-minima appears between 24 and 32 Hz for the voltages studied, in good agreement with the displacement measurements that peak for low frequencies. For frequencies above the minimum, the sound pressure level increases continuously and gradually until reaching a maximum at 1498 $\pm$ 6 [Hz]. The curves shown in Figure~\ref{Fig:SpeakerResponseAlone} are typical of a flat response of a real speaker. In anechoic conditions at the highest voltage, the speaker produces a maximum sound pressure of 115 dB at 30 cm from the dust cap.

\subsection{Frequency response of the cavity}

\begin{figure*}
\centering
\includegraphics[width=1.98\columnwidth]{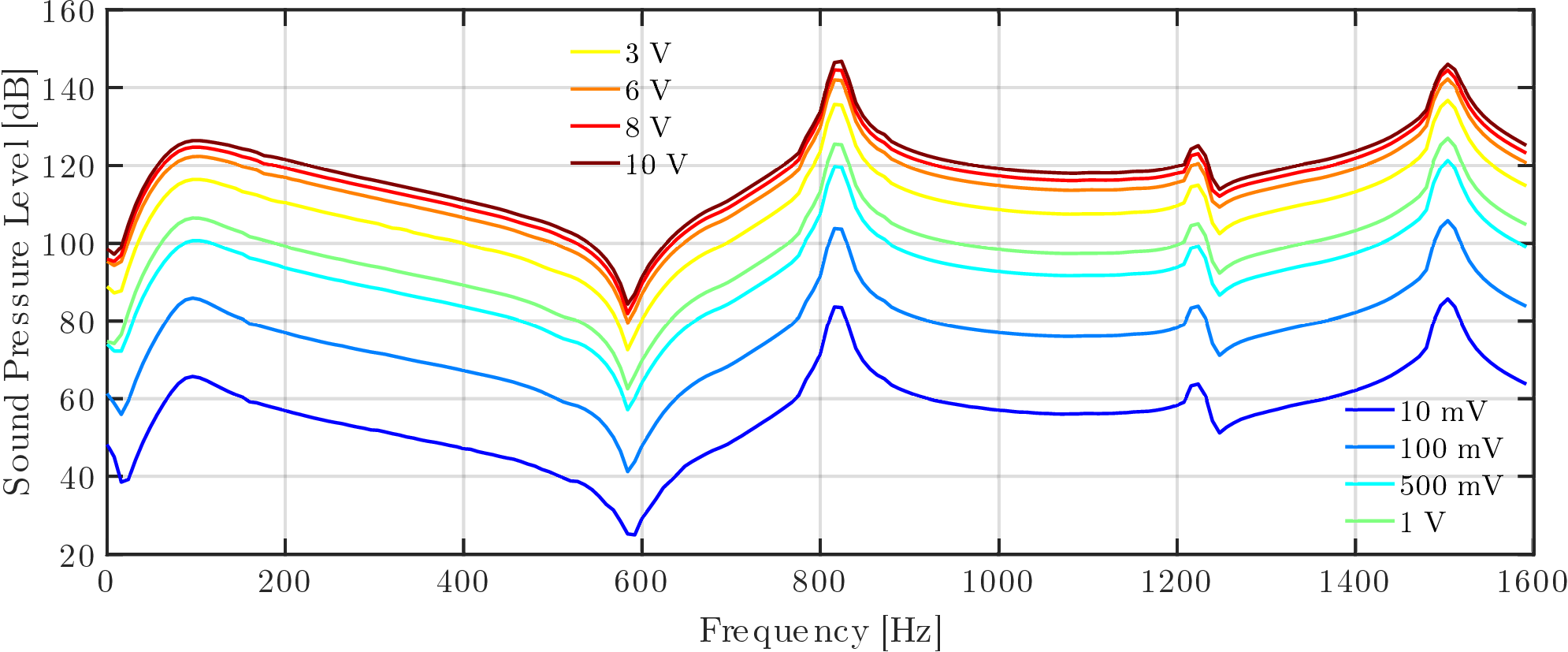}
\caption{(Color online) Cavity frequency response drive by the speaker. Maximum sound pressure level during a sine sweep measured in an anechoic chamber from 1 to 1600 Hz for different voltages.}
\label{Fig:CavityResponse}
\end{figure*}

The cavity of the SJ actuator is cubic with an inner side of 20 cm. The speaker attached to the wall seals the cavity, and air may only go in and out through the neck of the resonator. We tested the frequency response of the cavity by connecting the speaker to the audio amplifier while running a sine sweep from 1 to 1600 Hz at different voltages. The cavity response shown in Figure~\ref{Fig:CavityResponse} shows that the cavity exhibits characteristics similar to rooms. For example, a spike consistently appears at 817 $\pm$  3 Hz for the voltages studied. Such a spike corresponds to the first normal room mode, predicted at
\begin{equation}
	f = \frac{c_0}{\lambda_1} = \frac{c_0}{2 L},
	\label{Eq:1stNormalMode}
\end{equation}
where $c_0$ = 333 m/s is the local speed of sound estimated in the lab, $\lambda_1 = 2 L = 40 $ cm is the first wavelength that fits the cavity of the SJ actuator. Such a small calculation leads to a peak frequency of 833 Hz. One of the reasons the calculation does not conform to the measurement is the cone recess, meaning there is a longer average distance between the dust cap and the wall than the length of the wall.  We estimated that an additional length of 37 mm would be enough to match both frequencies, consistent with the length associated with the recessed volume. Two more spikes appear at 1232 and 1504 Hz and correspond to the first tangential and oblique modes. Calculations to estimate these modes can be performed similarly, as in Equation~\ref{Eq:1stNormalMode}. Despite peaks of similar magnitude, the third symmetric mode exhibits a higher level than the first for the lowest voltage, which inverts as voltage increases. We speculate that the maximum frequency of the speaker is related to this result. Linear room acoustics predicts the optimum frequency of the three symmetric spikes. However, Figure~\ref{Fig:SpeakerResponseAlone} shows another spike at lower frequencies. This spike is not symmetric as it develops, reaches its maximum on tenths of Hertz, and gradually reduces during hundreds of Hertz with its maximum at 103 $\pm$ 2 Hz. Such behavior results in a shape more similar to a hump than a spike, and its maximum value is about one order of magnitude ($\approx 20$ dB) below the dominating room mode. The maximum frequency of this spike does not correspond to any standard room mode.

\begin{figure*}
\centering
\includegraphics[width=.99\columnwidth]{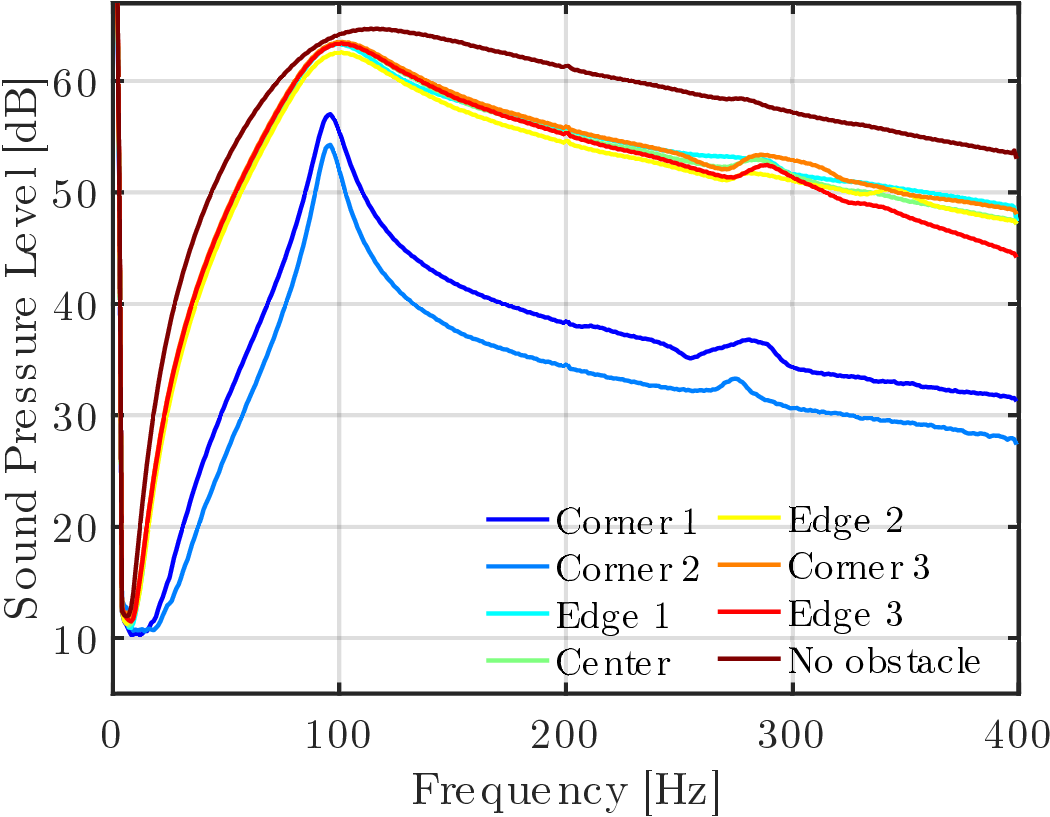}
\caption{(Color online) Sound pressure curves when an obstacle is placed in different positions of the cavity at 100 mV.}
\label{Fig:ObstacleResponse}
\end{figure*}

At low sound intensities, the behavior of the room mode depicted in Figure~\ref{Fig:CavityResponse} was strongly affected by placing an obstacle in different positions within the cavity. The obstacle was a cylindrical bar of 5.5 cm in diameter and 9 mm in height. Figure~\ref{Fig:ObstacleResponse} shows the result of this test, where we measured the sound pressure level within the cavity as a function of the frequency for an intensity of 100 mV. The color curves correspond to different positions of an obstacle within the cavity. The position labels in Figure~\ref{Fig:SetupSketch}(b) correspond to the floor wall. The maroon curve corresponds to the sound pressure level without an obstacle and is the highest of the curves. The maximum intensity decreased by more than 10 dB when we positioned the bar in Corner 2. 

Similarly, the intensity decreased more than 7 dB when we placed the obstacle in Corner 1. When we placed the obstacle on the edges, the center, and Corner 3, there was a slum decrease in the intensity but a significant change in the dispersion. The curves for Corner 1 and 2 show a sharper peak in the maximum frequency of the spectrum. This behavior was not observed for higher voltages, as all the curves took the shape of the maroon curve in Figure~\ref{Fig:ObstacleResponse}.

\subsection{Helmholtz Resonance}
\begin{figure*}
\centering
\includegraphics[width=1.98\columnwidth]{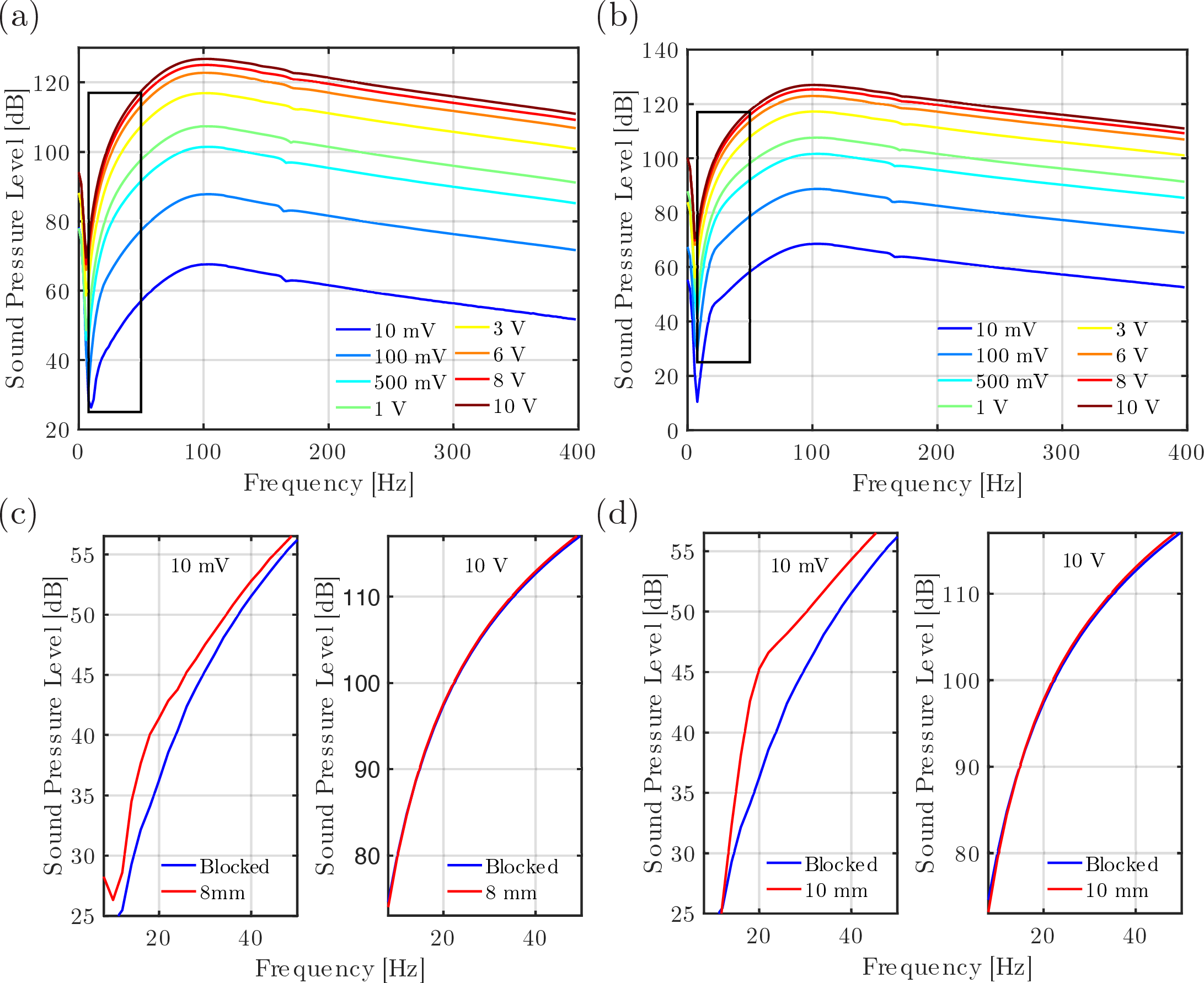}
\caption{(Color online) Sound pressure level measured using frequency analysis of the cavity open with a nozzle of (a \& c) 8 mm and (b \& d) 10 mm. (c \& d) show a close-up for the lowest and the highest voltages.}
\label{Fig:PresRespVSHelmholtz}
\end{figure*}

Once the nozzle is open to the ambient, a Helmholtz resonator couples to the SJ actuator. Here, we tested two nozzles with the same length and different diameters. Coupling another resonator may add to the complexity of the system. However, these DHOs couple in specific frequencies that can be estimated theoretically as 
\begin{equation}
	f_{He} = \frac{c_0}{2 \pi}\sqrt{\frac{A_n}{V_0 l_n}},
\end{equation}
where  $A_n$ is the cross-sectional area of the nozzle, $V_0$ is the cavity volume of the SJ actuator, and  $l_n$ is the length of the nozzle. This calculation leads to Helmholtz frequencies of 20 and 25 Hz for the 8 and 10 mm nozzle diameters, respectively. Figure~\ref{Fig:PresRespVSHelmholtz}(a \& b) shows the sound pressure level spectra when the SJ actuator is open using the 8- and 10-mm nozzles. These curves are, in essence, the same as the ones in Figure 4 up to 400 Hz and the maroon curve without obstacle in Figure 5. Despite the similarity, there are differences with the Helmholtz frequencies calculated previously. Figure~\ref{Fig:PresRespVSHelmholtz} (c \& d) shows a close-up of the regions marked with black rectangles in Figure~\ref{Fig:PresRespVSHelmholtz} (a \& b). Figure~\ref{Fig:PresRespVSHelmholtz} (c) shows the effect of an 8 mm nozzle on the lower (left) and the higher (right) voltages. Figure~\ref{Fig:PresRespVSHelmholtz} (d) shows the result of a 10 mm nozzle on the lower (left) and the higher (right) voltages. In this figure, Helmholtz resonators add a minor effect to the spectrum, even for the lower voltages. The effect becomes negligible for voltages higher than 100 mV, compared with the spectra measured when blocked. Thus, it becomes clear that the nonlinear room mode developing within the cavity is the dominant effect leading to the sound pressure level.

\begin{figure*}
\centering
\includegraphics[width=1.98\columnwidth]{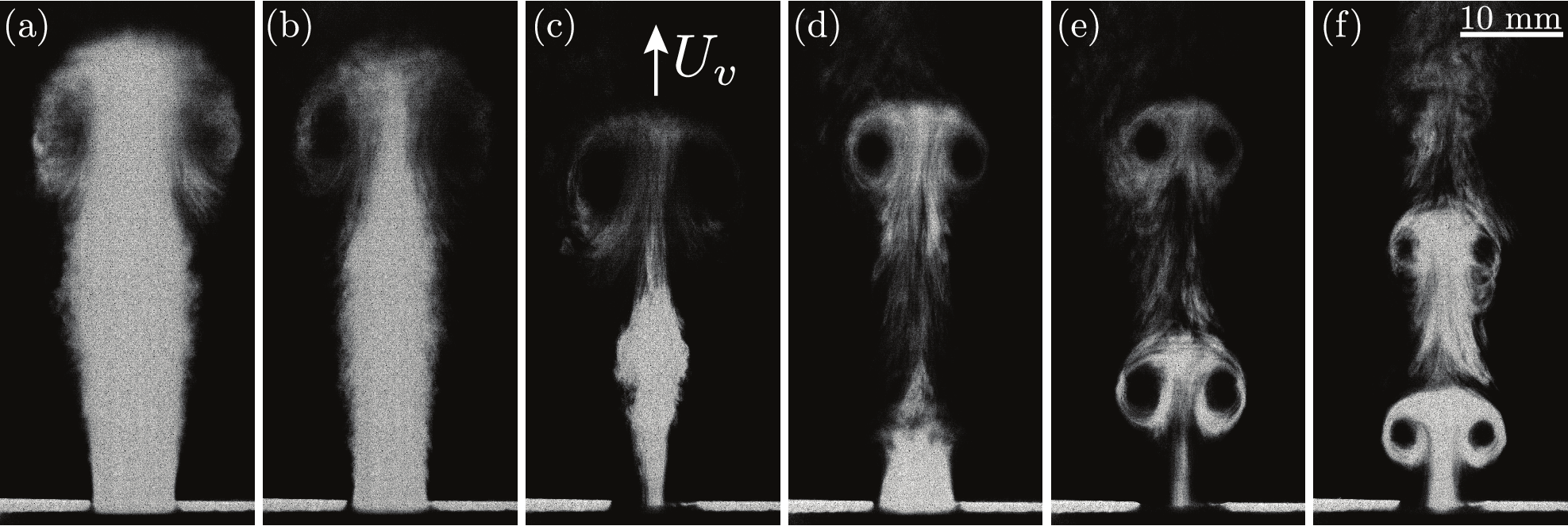}
\caption{(Color online) Jets and vortex rings visualized at 8 V. Visualization experiments at driving frequencies of (a) 40 Hz, (b) 80 Hz, (c) 110 Hz, (d) 130 Hz, (e) 140 Hz, (f) 150 Hz.}
\label{Fig:StrokeLengthPhotos}
\end{figure*}

\begin{figure*}
\centering
\includegraphics[width=1.98\columnwidth]{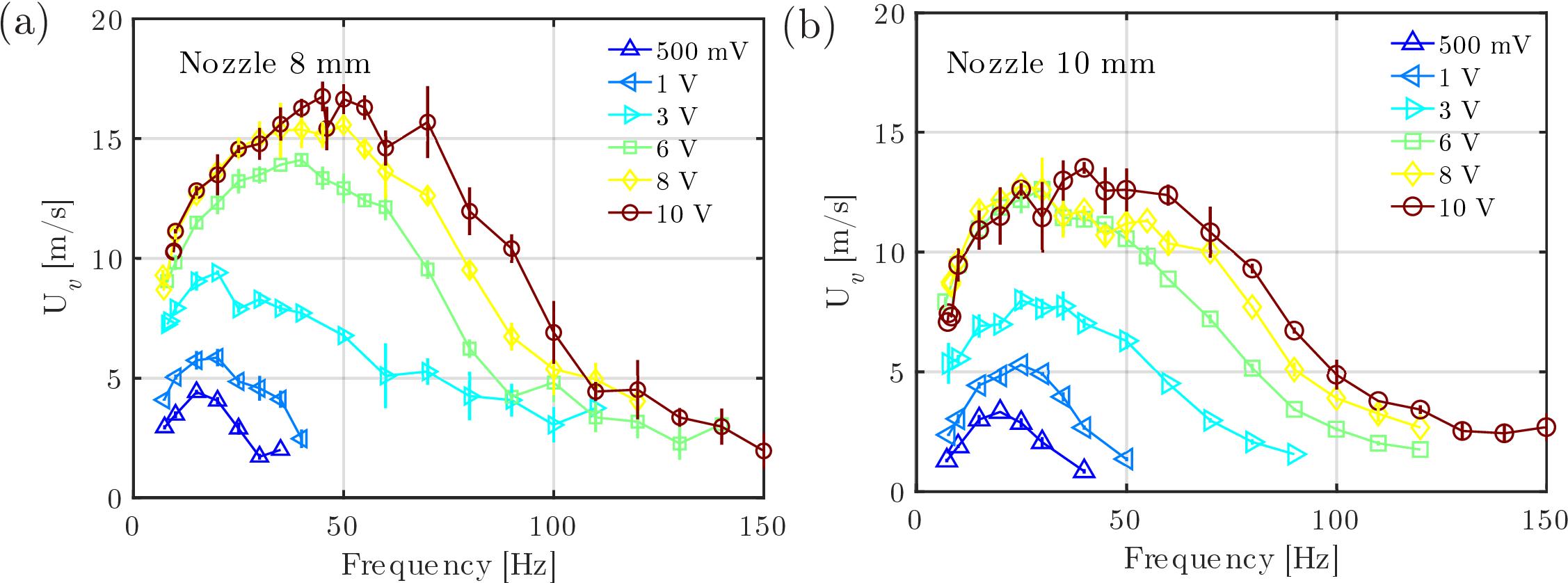}
\caption{(Color online) Velocity measurements of vortices expelled from the the SJ actuator at different frequencies and voltages. Two nozzles were used: (a). The diameter is 8 mm. (b). The diameter is 10 mm.}
\label{Fig:VortexVelocity}
\end{figure*}

Commonly, the velocity of SJ mirrors the sound pressure curves. To test the effect of the nonlinear room mode on the jet, we measured the jet velocity for different conditions of voltage and frequency. To perform the velocity measurements, we visualized the SJ flow at the exit of the SJ actuator. Figure~\ref{Fig:StrokeLengthPhotos} shows a panel where the frequency is increased from left to right while keeping the voltage constant at 8 V.  The arrow in Figure~\ref{Fig:StrokeLengthPhotos}(c) depicts the velocity direction. Figure~\ref{Fig:VortexVelocity} shows systematic velocity measurements as a function of the frequency for voltages from 500 mV to 10 V. Smaller voltages did not produce flow. The vertical lines of Figure~\ref{Fig:VortexVelocity} represent error bars estimated as the standard deviation of the velocity of at least ten measurements. We observed that the velocity increases with frequency, reaching a maximum at intermediate frequencies. For lower voltages, 500 mV and 1 V, the maximum velocity is located at about 20 Hz. However, for higher voltages, the maximum velocity shifts toward higher frequencies. This behavior is similar when we compare two nozzles of different diameters.

While a change of a diameter from 8 to 10 mm represents only a 25\% increase, the area of the cross-section is increased by more than 50\%. The increased area balances with the jet velocities decrease. The maximum velocities measured in Figure~\ref{Fig:VortexVelocity} (a) are around 17 m/s, while such a velocity is reduced to 12 m/s when the nozzle has a larger aperture, implying the flow rate keeps constant.

Each velocity curve in Figure~\ref{Fig:VortexVelocity} shows a maximum velocity at a given frequency. The optimum frequency is voltage-dependent, shifting toward higher frequencies for higher voltages. In Figure~\ref{Fig:VortexVelocity}(a), the maximum frequency for 500 mV is about 20 Hz, and it gets closer to 50 Hz for 10 V. We observed a similar behavior for the lower voltages in Figure~\ref{Fig:VortexVelocity}(b). The velocity does not exhibit a clear peak for higher voltages, and the maximum velocities are observed for 30 to 50 Hz if accounting for the error bar..

\section{Discussion}
The reason for the frequency shift from 20 to about 50 Hz documented in Figure~\ref{Fig:VortexVelocity}  is not self-explanatory from the data presented up to this point. When conditions lead to SJ formation, the maximum frequency at lower voltages is misleadingly close to the Helmholtz frequency of the resonator. Despite using two nozzles of different diameters, their resonating frequencies coincided with the minimum response of the speaker. This condition limited the contributions of the Helmholtz resonance from the beginning. Helmholtz resonance only impacts the sound pressure for the lower voltages, 10 and 100 mV. Our SJ actuator only expells vortices when the voltage is at least 500 mV. We did not measure any significant impact of the Helmholtz resonance on the sound pressure at 500 mV nor at 1 V. The SJ regime of this device is far from the regime where the Helmholtz resonance plays any role in the pressurized cavity.

\begin{figure*}
\centering
\includegraphics[width=1.98\columnwidth]{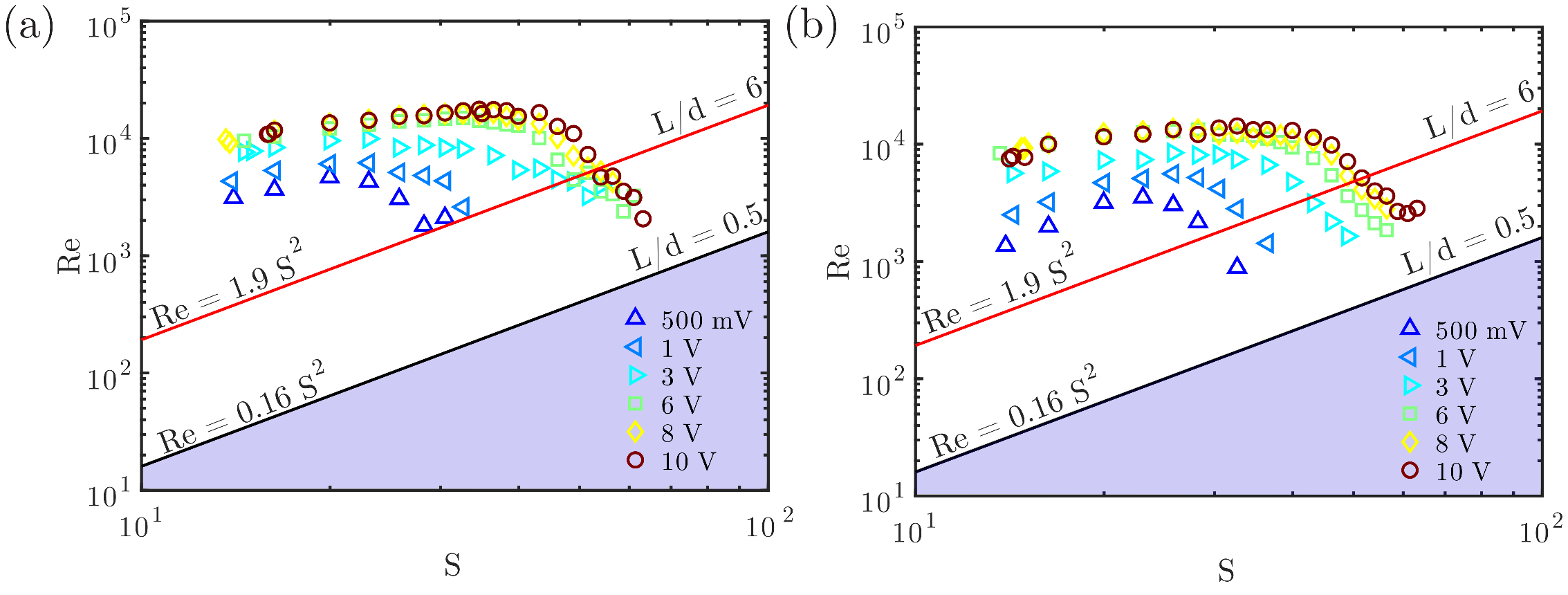}
\caption{(Color online) Stokes vs Reynolds transition diagram. (a). The diameter is (a) 8 mm and (b) 10 mm.}
\label{Fig:StokesVSReynolds}
\end{figure*}

Furthermore, even at the two lower voltages, the Helmholtz resonance only represents a small addition to the sound pressure within the cavity, as seen in Figures~\ref{Fig:PresRespVSHelmholtz}(c \& d), meaning the asymmetric mode is overwhelmingly dominant even here. However, one more criterion plays a role in the performance of the SJ actuator: the vortex formation criteria. This criterion depends on two dimensionless parameters, the Stokes and the Reynolds number. The Stokes number is defined as 
\begin{equation}
	\textrm{S} = d\sqrt{\frac{2 \pi f}{\nu}},
\end{equation}
where $d$ is the diameter of the nozzle, $f$ is the driving frequency, and $\nu$ is the kinematic viscosity. For an SJ driven by a sinusoidal waveform, a modified Reynolds number is defined as
\begin{equation}
	\textrm{Re} = \frac{2 U_v d}{\nu}.
\end{equation}
The numerical prefactor on the definition of the Reynolds number corresponds to the velocity integration during the time of discharge, {\it i.e.}, half the cycle of a sine wave. The formation criteria\cite{Brouvckova2016} state that vortices may only form if $\textrm{Re} > 0.16\; \textrm{S}^2$. Figure~\ref{Fig:StokesVSReynolds} depicts this space parameter, where the purple area corresponds to the region where vortices do not form. The stroke length may be introduced as an aid to interpreting Figure~\ref{Fig:StokesVSReynolds}. The stroke length is defined as $L = U_v/f$, and the dimensionless stroke length is defined as  in the context of SJs. Thus, the solid black line in Figure~\ref{Fig:StokesVSReynolds} corresponds not only to the formation criteria but may also be interpreted as a dimensionless stroke length of $L/d = 0.5$. In this device, vortices do not stop forming at a constant dimensionless stroke length. Despite curve $\textrm{Re} = 0.16\; \textrm{S}^2$ representing a transition from flow to no-flow, crossing through different curves above signifies a propulsion decrease. When the 8 mm nozzle is attached, voltages of 500 mV and 1 V vortices stop forming at $L/d <6$, as seen in Figure~\ref{Fig:StokesVSReynolds}(a). But, for higher voltages, vortices still form for shorter stoke lengths, approaching the formation criteria. Figure~\ref{Fig:StokesVSReynolds}(b) shows a similar behavior for the nozzle of 10 mm. The main difference between Figure~\ref{Fig:StokesVSReynolds}(a \& b) is that vortices still form at stroke lengths smaller than $L/d <6$ for the lower voltages. The red lines in Figure~\ref{Fig:StokesVSReynolds} correspond to the same boundary given by the dimensionless stroke length. However, for the nozzle of 10 mm, vortices form between the black and the red line, and the same does not occur for the 8 mm nozzle. This behavior is consistent with the transitional area reported by Zhou et al.\cite{Zhou2009} (2009), although not defined on the same boundaries.

Now, we elaborate on the conditions needed for an SJ actuator to produce flow at a single dimensionless stroke length. First, we may assume that for a laminar regime, the pressure drop between the cavity and the environment scales linearly with speed during discharge, as $U_v \sim \Delta P$. The complete scaling is known as the Hagen-Poiseuille equation. This scaling is valid for Reynolds numbers up to 3000. As the flow transitions to turbulent, the linear relation gradually shifts to $U_v \sim \Delta P^{1/2}$. The rate of such transition depends on the relative roughness of the pipe. Fixing the dimensionless stroke length in the Stokes-Reynolds chart means the relation between velocity and frequency should be constant, translating to $U_v \sim f$ for a given nozzle diameter. Thus, to preserve the stroke length, the pressure should increase linearly with the frequency, $f \sim \Delta P$, when the flow is laminar. This dependency is weaker when a turbulent jet fully develops, $f \sim \Delta P^{1/2}$. 

\begin{figure}
\centering
\includegraphics[width=0.99\columnwidth]{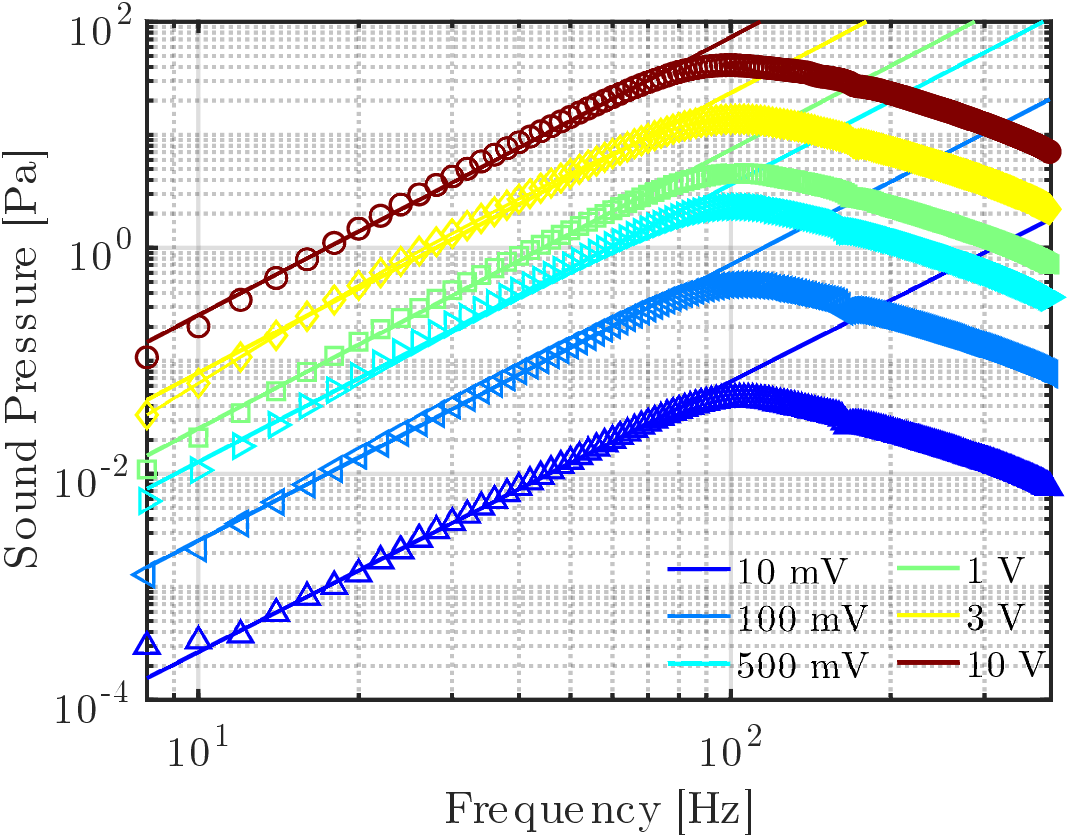}
\caption{(Color online). Sound pressure vs frequency in log-log scale. The solid lines correspond to the a power-law fit  of the growth rate to estimate the exponent.}
\label{Fig:ModeGrowthRate}
\end{figure}

Unlike previous investigations on SJ actuators, the external flow is driven by the pressure accumulated within the cavity by the non-linear resonant mode. Hereunder, we present an argument to scale the maximum pressure growth produced by the resonant mode to drive the flow at higher frequencies. Figure~\ref{Fig:ModeGrowthRate}(a) shows the sound pressure as a function of the frequency for different voltages in the log-log scale. The pressure accumulation in the cavity due to the development of the resonant mode during low frequencies may be simplified using a power law of the form
\begin{equation}
	\Delta P = a f^n,
\end{equation}
where $a$ is a prefactor that is a function of the voltage and $n$ is the exponent. The solid lines in Figure~\ref{Fig:ModeGrowthRate}(a) correspond to a power law fit. The prefactor $a$ varies continuously as a function of the voltage, as seen in Table~\ref{Tab:PowerLawFit}. However, the exponent is relatively the same, with a mean value of $\bar{n} = 2.46\pm0.03$. For simplicity, we may assume that the exponent is $n = 5/2$. According to the power-law-fit, the frequency scales with the pressure as $f\sim\Delta P^{2/5}$. This scaling has a smaller exponent than needed to keep a constant dimensionless stroke length, even for a fully developed turbulent jet. Even at its highest growth rate, the pressure developed in the cavity by the resonant mode is insufficient to maintain a constant stroke, thus resulting in the jet weakening and ultimately inhibiting its formation. The experimental points in Figure~\ref{Fig:StokesVSReynolds} exhibit the insufficient growth rate as a lower exponent than a fixed dimensionless stroke length. The direction change at higher Stokes numbers corresponds to the exponent decay of the pressure curve. Lower voltages result in shorter initial stroke lengths, being more sensitive to the stroke length drop for higher frequencies. The combination of the asymmetric resonant mode and the formation criteria results in the apparent frequency shift seen in Figure~\ref{Fig:StokesVSReynolds}. But, this behavior is not related to the Helmholtz resonance.

We suppose that the cubic shape plays a role in the development of the asymmetric room mode since the three length scales related to the size are equally important, which contrasts with the formation of resonant modes on closed and narrow ducts, where the length of the duct is much more significant than its cross-section.

\begin{table}
\centering
\caption{\label{Tab:PowerLawFit}{Pre-factors and exponents fitted to the maximum growth rate of the asymmetric room mode to a power law.}}
\begin{tabular}{|c|c|c|} \hline
V [mV] & $a$[Pa] & $n [1]$  \\  \hline
1$\times$10$^1$ & 1.0737$\times$10$^{-6}$ & 2.3919\\
1$\times$10$^2$ & 9.2743$\times$10$^{-6}$ & 2.4406\\
5$\times$10$^2$ & 4.4698$\times$10$^{-5}$ & 2.4539\\
1$\times$10$^3$ & 8.5289$\times$10$^{-5}$ & 2.4685\\
3$\times$10$^3$ & 2.6217$\times$10$^{-4}$ & 2.4750\\
6$\times$10$^3$ & 5.0276$\times$10$^{-4}$ & 2.4876\\
8$\times$10$^3$ & 6.8815$\times$10$^{-4}$ & 2.4732\\
1$\times$10$^4$ & 8.7928$\times$10$^{-4}$ & 2.4592\\ \hline
\end{tabular}
\end{table}

\section{Conclusions}
We studied an SJ actuator driven by a speaker whose cavity operates in a compressible regime. The different elements of the SJ actuator are studied independently. Despite being a peak on its electric impedance, this does not affect its mechanical or acoustic response. We tested the mechanical response of the speaker using accelerometry to estimate the RMS displacement at different voltages. The speaker showed large displacements for low frequencies that gradually diminished as the frequency increased. We tested the acoustic response by driving a sinusoidal frequency sweep at a constant voltage and performing frequency analysis. The speaker exhibited a flat response up to 1600 Hz. The element of the SJ actuator that dominated its dynamics was its large cavity.

The cavity volume is more than one order of magnitude larger in previous reports. Such a volume is significantly larger than other cavities. But, it is small compared with standard rooms. A typical theory implemented on SJ actuators assumes the wavelengths are much larger than the device. Therefore, their field properties collapse to their average values, and their elements are simplified. Such an assumption is unused in standard rooms, where reflection and diffraction form standing waves that result in room modes. A nonlinear room mode develops within the cavity of our SJ actuator at very low frequencies. The spectrum of this room mode is different from others, as its frequency distribution is significantly asymmetric, and its maximum is one order of magnitude weaker. Despite this fact, this resonant mode dominates the dynamics, above the speaker and Helmholtz resonance. It may underlie that despite the size of the cavity being considerably smaller than the wavelength, it is not negligible anymore. We suppose the asymmetric room mode results from a nonlinear effect within the cavity. The room-mode intensity can diminish for low voltages, and its frequency distribution is focused when adding an obstacle in different cavity positions. It may point to the formation of nodes and antinodes, as in standard room acoustics. The velocity measurements in Figure~\ref{Fig:VortexVelocity} show an apparent resonance shift to higher frequencies as voltage increases. We showed the shift results from shorter stroke lengths as frequency increases due to the driving mechanism that produces vortices and jets. We believe this room mode could be included as part of the design of SJ actuators to generate a flatter resonant response around resonance than other devices.

\section*{Acknowledgements}
This work was supported by UNAM-PAPIIT IN107024. J.F. Hern\'andez-S\'anchez and S. S\'anchez acknowledge the CONAHCYT (Mexico) to finance the project CF-2023-I-1344. Benjamin Valera Orozco, Antonio P\'erez-L\'opez, and Ricardo Dorantes-Escamilla provided technical support for the experimental work. The authors thank Pablo Rend\'on for their support and insightful discussions on this project. 

\section*{Data Availability Statement}
The data that support the findings of this study are available from the corresponding author upon reasonable request.

\bibliographystyle{elsarticle-num-names}
\bibliography{Manuscript}
\end{document}